\documentclass[preprint,12pt]{elsarticle}

\usepackage{amsmath}

\journal{Physica A}

\begin{document}

\begin{frontmatter}
\title{Opinion dynamics as a movement in a bistable potential.}

\author{Piotr Nyczka, Jerzy Cis{\l}o, Katarzyna Sznajd-Weron}
\address{Institute of Theoretical Physics, University of Wroc{\l}aw, pl. Maxa
Borna 9, 50-204 Wroc{\l}aw, Poland }

\begin{abstract}
In this paper we investigate the model of opinion dynamics with anticonformity on a complete graph. We show that below some threshold value of anticonformal behavior spontaneous reorientations occur between two stable states. Dealing with a complete graph allows us also for analytical treatment. We show that opinion dynamics can be understood as a movement of a public opinion in a symmetric bistable effective potential. We focus also on the spontaneous transitions between stable states and show that a typical waiting time can be observed. 
\end{abstract}


\end{frontmatter}
\section{Introduction}
The word "revolution" came from the Latin "revolutio", which means "a turn around" -- a fundamental change that takes place in relatively short period of time. Although, revolutions occurred many times in human history, scientists still investigate what are possible reasons for a revolution. Particularly interesting empirical examples are recent protests in Egypt and Tunisia. In Egypt protests started on Tuesday, January 25 2011. These were the first protests on such a large scale to be seen in Egypt since the 1970s. What was the reason, why these protests begun on January 25? Sociologists claim that it was a domino effect -- Egyptian revolution has been inspired by the successful revolution in Tunisia. Immediately the next question arise -- what was the reason for the revolution in  Tunisia? Again, sociologists and journalists claim that the answer is known  -- it has been sparked by the suicide of a young man who could not find a job and was barred from selling fruit without a permit. There are of course many other examples of revolutions that took place in history. Usually, reasonable explanation is given by sociologists \emph{post factum}. On the other hand, knowing the history one can always find some explanation. The problem is that we can hardly predict exact moment or place of a revolution. Why? Is it true that every overturn has a direct source? Physicist's experience shows that if a system supports two stable states, then fluctuations can cause random transitions between those two states \cite{GHJM2009,HTB1990}. Recently such a spontaneous transitions, induced by a random noise, have been shown in a one-dimensional Sznajd model of opinion dynamics \cite{KSW2010}. 

It should be noticed that the idea to introduce a random noise to the Sznajd model has been proposed already in the original paper from 2000 \cite{SWS2000}. However, in spite of the idea no systematic studies have been performed. In 2005 de la Lama et al. have included a social temperature effect in the model \cite{LLW2005}. They have assumed that at every time step an agent followed the rules of the Sznajd model with probability $p$, while with a probability $1-p$ rules were not fulfilled (an agent adopted the opposite option than the one dictated by the rules) \cite{LLW2005}. Such a modification has led to so called contrarians behavior \cite{G2004}. It has been shown that for $p>p_c$ the system was bistable with a probability density of magnetization having two maxima at $m=m_{\pm}$ \cite{LLW2005}. It has been claimed that for $p>p_c$ the system got ordered spontaneously selecting one of the stable solutions $m=m_{\pm}$, while for $p<p_c$ the system became monostable and disordered \cite{LLW2005}. Results obtained recently in \cite{KSW2010} for one-dimensional system confirmed those obtained by de la Lama et al. Moreover, it has been shown that the finite system with random 'contrarian-like' noise led to spontaneous transitions between two ordered stable states $m=m_{\pm}=\pm 1$ \cite{KSW2010}.

It should be recall here that the possibility of spontaneous transitions in the Sznajd model with noise has been also investigated by Wio et al. by including an external modulated field. This is well known that usually bistable systems response to a periodic external force in the presence of noise. This phenomena is known as a stochastic resonance and it has been widely studied in various systems over last 30 years \cite{GHJM2009,GHJM1998}. Wio et al. have shown that the system described by the Sznajd model with contrarians did not show conventional stochastic resonance, but so called system size stochastic resonance. 

In this paper we investigate different problem -- spontaneous transitions not induced by any external force. We consider again generalized version of the Sznajd model introduced in \cite{KSW2010} in which two types of social response (conformity and anticonformity) are possible. In \cite{KSW2010} we have shown, performing Monte Carlo simulations for one-dimensional system, that for small values of anticonformity level, spontaneous reorientations occur between two opposite consensus states ('all spins up' or 'all spins down'). In this paper we consider the same model on a complete graph, which allows us for analytical treatment. Moreover, we show that results obtained for the complete graph are qualitatively different from those obtained for the one-dimensional system and more suitable to describe opinion dynamics in real social systems. 

We would like to clarify here that in econophysics and economy term 'anticonformity' refers usually to contrarian behavior -- a preference for taking a position opposed to that of the majority. In \cite{KSW2010} we have used a term 'anticonformity', that is usually used in social psychology \cite{Myers_1996,Nail_2000}, to distinguish between two commonly mixed up types of social response - 'anticonformity' and 'nonconformity'. While anticonformity is similar to conformity in the sense that both (conformers and anticonformers) acknowledge the group norm (the conformers agree with the norm, the anticonformers disagree), nonconformity means independence. As long as we do not focus on particular applications of the model (financial, social or any others) we can use interchangeably both terms 'contrarian' and 'anticonformists'.

\section{The model}
We consider a set of $N$ individuals, which are described by the binary variables $S=1$ ($\uparrow$) or $S=-1$ ($\downarrow$) on a complete graph. At each elementary time step two individuals (denoted by $\uparrow$ or $\downarrow$) are chosen at random, and they influence a third randomly chosen individual (denoted by $\Uparrow$ or $\Downarrow$) in the following way:
\begin{eqnarray}
\uparrow\uparrow\Downarrow \rightarrow \uparrow\uparrow\Uparrow & \mbox{ with probability } p_1, & \mbox{no change with }  1-p_1\nonumber \\
\downarrow\downarrow\Uparrow \rightarrow \downarrow\downarrow\Downarrow & \mbox{ with probability } p_1, & \mbox{no change with }  1-p_1 \nonumber\\
\uparrow\uparrow\Uparrow \rightarrow \uparrow\uparrow\Downarrow & \mbox{ with probability } p_2 , & \mbox{no change with }  1-p_2\nonumber \\
\downarrow\downarrow\Downarrow \rightarrow \downarrow\downarrow\Uparrow & \mbox{ with probability } p_2, & \mbox{no change with }  1-p_2.
\label{def_mod}
\end{eqnarray}
First two processes correspond to conformity, and the next two describe anticonformity. Above model is a particular case of the generalized Sznajd model introduced in \cite{Kondrat_2009}. It has been suggested in \cite{Kondrat_2009} that significant parameter of the model is the ratio $r=p_2/p_1$ and changing absolute values of $p_1$ and $p_2$ influence only a time scale. Moreover, in the next section we show analytically that indeed stationary states depend only on the ratio $r$. In the previous paper \cite{KSW2010} we have decided to investigate the case in which $p_2=1$ and $p_1 \in (0,1) $ is the only parameter of the model. In this paper we keep this assumption for Monte Carlo simulations, although we present also general analytic results for arbitrary values of $p_1,p_2$. 

In our previous paper \cite{KSW2010}, to investigate the model, we have provided Monte Carlo simulations with the random sequential updating mode and thus the time $t$ has been measured in the Monte Carlo Sweeps (MCS) which consisted of $N$ elementary updatings, i.e. $\Delta_t=1/N$. As usually, we have defined a public opinion as a magnetization of the system:
\begin{equation}
m(t) = \frac{1}{N} \sum_{k=1}^N S_k(t).
\label{eq:m}
\end{equation}
Monte Carlo results for the one-dimensional system have shown that for small values of anticonformity level $p_2$ spontaneous reorientations occur between two extreme values of $m = \pm 1$. We have shown that there is a special value of anticonformity level $p_2=p^*$ below which the system stays for most of time in the one of two possible consensus states ($m = \pm 1$) and spontaneous reorientations occur randomly. Above the threshold, i.e. for $p_2>p^*$ system becomes monostable with magnetization fluctuating around zero \cite{KSW2010}.  

In this paper we investigate the model on a complete graph, which allows us for analytical treatment. Moreover, we show that in case of a complete graph, spontaneous reorientations occur not between two extreme values of $m = \pm 1$, but between two stable solutions $m = m_{\pm}$ for $p_2/p_1=r<r^*=1/3$. In other words, in the case of a complete graph we observe continuous phase transition at $r^*$, while discontinuous transition in a case of one-dimensional lattice.  

In this paper we focus mostly on the spontaneous transitions between stable states, which were not investigated analytically up till now. Such a random transitions between two stable states are particularly important from social perspective to understand revolutions, protests and other fundamental social changes that take places in relatively short period of time.
 
\section{Analytical results for the infinite system}
In the case of a complete graph the state of the system is completely described by the magnetization $m$ defined by Eq. (\ref{eq:m}) and takes the following values:
\begin{equation}
m=\left\{ -1,-1+\Delta_N,-1+2\Delta_N, \ldots, 1 \right\},
\label{eq_state_mag}
\end{equation}
where $\Delta_N=2/N$.

\begin{figure}
\begin{center}
\includegraphics[width=12cm]{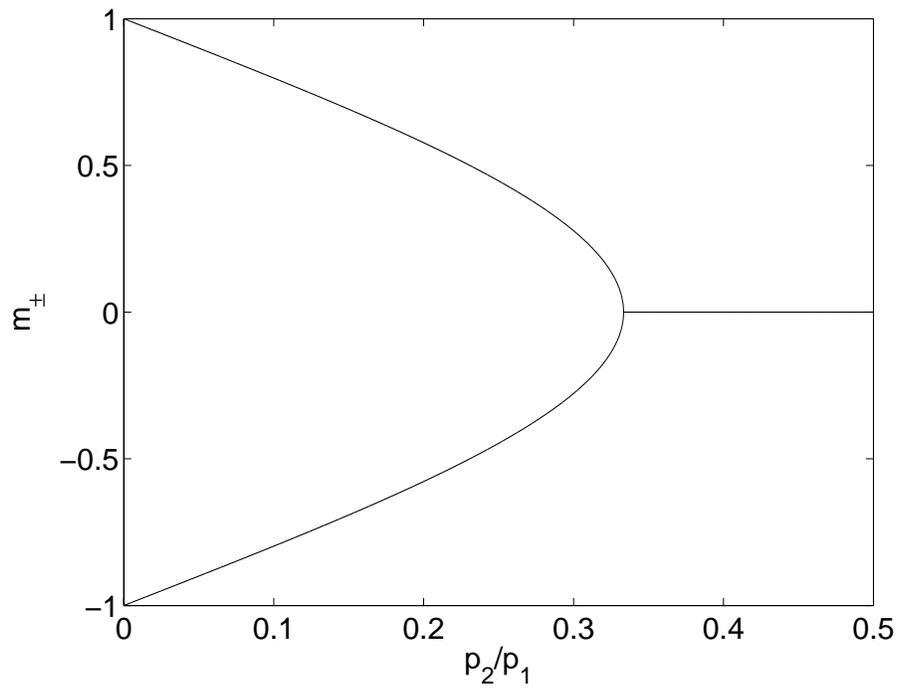}
\caption{Stationary values of magnetization $m_{\pm}$ as a function of the ratio $r=p_2/p_1$. Below the threshold value $r=1/3$ there are two possible values of magnetization, above the threshold there is no majority in the system. From social perspective the system is in a stalemate state.}
\label{fig_analit}
\end{center}
\end{figure}

\begin{figure}
\begin{center}
\includegraphics[width=12cm]{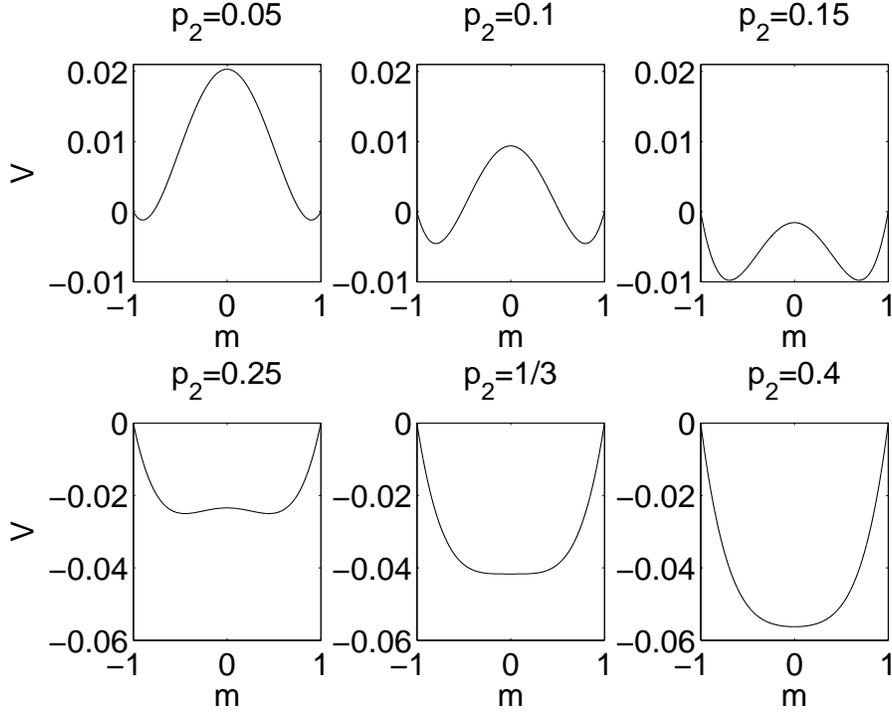}
\caption{Effective potential $V$ as a function of magnetization for $p_1=1$. As seen below threshold value $p_2=1/3$ potential is bistable -- there are two stable states at $m=m_{\pm}$}
\label{pot}
\end{center}
\end{figure}

Let us denote by $N_{\uparrow}$ number of spins 'up' and by  $N_{\downarrow}$ number of spins down:
\begin{eqnarray}
N_{\uparrow}+N_{\downarrow}& = & N \nonumber\\
N_{\uparrow}-N_{\downarrow}& = & mN.
\label{eq_NpNm}
\end{eqnarray} 
From above equations we can easily derive the formula for the probability of choosing randomly spin 'up' (concentration $c$ of 'up' spins) as:
\begin{eqnarray}
c & = & \frac{N_{\uparrow}}{N} = \frac{1+m}{2} \equiv c. 
\label{mag_p}
\end{eqnarray} 
We can now calculate the probability $p_+$ that the concentration of 'up' spins increases $c \rightarrow c+1/N$ ($m \rightarrow m+\Delta_N$) and the probability $p_-$ that $c$ decreases $c \rightarrow c-1/N$ $m \rightarrow m-\Delta_N$).
As seen from the definition of the model given by (\ref{def_mod}) only $4$ of $8$ possible spin configurations are active $\uparrow\uparrow\downarrow,\uparrow\uparrow\uparrow,\downarrow\downarrow\uparrow,\downarrow\downarrow\downarrow$. If an inactive configuration is chosen, nothing happens. In the case of the infinite system, probabilities of choosing active configurations are the following:
\begin{eqnarray}
p_{\uparrow\uparrow\downarrow} & = & c^2(1-c), \nonumber \\
p_{\uparrow\uparrow\uparrow} & = &  c^3, \nonumber \\
p_{\downarrow\downarrow\uparrow} & = &  (1-c)^2c, \nonumber \\
p_{\downarrow\downarrow\downarrow} & = & (1-c)^3. 
\label{eq_prob_inf}
\end{eqnarray}
From (\ref{def_mod}) transition probabilities can be calculated:
\begin{eqnarray}
p_{\uparrow\uparrow\downarrow \rightarrow \uparrow\uparrow\uparrow} & = &  p_1 \cdotp p_{\uparrow\uparrow\downarrow},   \nonumber \\
p_{\uparrow\uparrow\uparrow \rightarrow \uparrow\uparrow\downarrow} & = & p_2 \cdotp p_{\uparrow\uparrow\uparrow}, \nonumber\\
p_{\downarrow\downarrow\uparrow \rightarrow \downarrow\downarrow\downarrow} & = & p_1 \cdotp p_{\downarrow\downarrow\uparrow}, \nonumber \\
p_{\downarrow\downarrow\downarrow \rightarrow \downarrow\downarrow\uparrow} & = & p_2 \cdotp p_{\downarrow\downarrow\downarrow}.
\label{eq:prob} 
\end{eqnarray}
Probabilities $p_+$ and $p_-$ that the concentration of 'up' spins increases/decreases are equal:
\begin{eqnarray}
p_+  & = & p_1 \cdotp p_{\uparrow\uparrow\downarrow} + p_2 \cdotp p_{\downarrow\downarrow\downarrow}, \nonumber \\
p_-  & = & p_1 \cdotp p_{\downarrow\downarrow\uparrow} + p_2 \cdotp p_{\uparrow\uparrow\uparrow},
\label{eq:prob0} 
\end{eqnarray}
and finally:
\begin{eqnarray}
p_+ & = & (1-c)^3p_2+c^2(1-c)p_1, \nonumber\\
p_- & = & c^3p_2 +c(1-c)^2p_1.
\end{eqnarray} 
After simple algebraic transformation we obtain the evolution equation: 
\begin{eqnarray}
c' & = & c + \frac{1}{N} p_+ - \frac{1}{N} p_- \nonumber\\
& = & c + \frac{1}{N} \left[ -2(p_1+p_2)c^3 + 3(p_1+p_2)c^2 -(p_1+3p_2)c + p_2 \right],
\label{evol_c}
\end{eqnarray}
where $c' \equiv c(t+\Delta_t)$ and $c \equiv c(t)$.  We can easily calculate the fixed points $c'=c$ of above equation:
\begin{eqnarray}
c_1 & = &  \frac{1}{2} \nonumber\\
c_2 & = &  \frac{1}{2}\left[1+\sqrt{\frac{p_1-3p_2}{p_1+p_2}}\;\right] = \frac{1}{2}(1+\sqrt{d})\nonumber\\
c_3 & = &  \frac{1}{2}\left[1-\sqrt{\frac{p_1-3p_2}{p_1+p_2}}\;\right] =  \frac{1}{2}(1-\sqrt{d}),
\label{c_st}
\end{eqnarray}
where useful quantity 
\begin{equation}
d = \frac{p_1-3p_2}{p_1+p_2}.
\label{def_d}
\end{equation}
For $d<0 \rightarrow r=p_1/p_2>1/3$ the only real solution of equation (\ref{evol_c}) is $c=c_1$. For $d \ge 0 \rightarrow  r=p_1/p_2 \le 1/3$ we have two stable solutions $c=c_2$ and $c=c_3$, while $c=c_1$ becomes unstable.
From (\ref{c_st}), using formula (\ref{mag_p}), we obtain immediately stationary values of magnetization in a very simple form (see Fig. \ref{fig_analit}):
\begin{eqnarray}
m_{st} & = & m_{\pm} = \pm \sqrt{d} \mbox{ for }  r<1/3 \nonumber \\ 
m_{st} & = & m_0 = 0 \mbox{ for } r \ge 1/3. 
\label{m_st}
\end{eqnarray}
It should be noticed that $F=p_+-p_-$ can be treated as an effective force -- $p_+$ drives the system to the state 'spins up', while $p_-$ to 'spins down':
\begin{eqnarray}
F=p_+-p_-=-2(p_1+p_2)c^3 + 3(p_1+p_2)c^2 -(p_1+3p_2)c + p_2. 
\label{effective_force}
\end{eqnarray}
Therefore we can easily calculate the effective potential:
\begin{eqnarray}
V=\frac{1}{2}(p_1+p_2)c^4 - (p_1+p_2)c^3 + \frac{1}{2}(p_1+3p_2)c^2 - p_2c.
\end{eqnarray}
Using formula (\ref{mag_p}) we obtain the dependence between effective potential $V$ and magnetization $m$. As seen from Fig.\ref{pot} below the threshold value $r=p_2/p_1 < 1/3$ potential is bistable -- there are two stable states at $m=m_{\pm}$ given by equation (\ref{m_st}). For $r \ge 1/3$ the only stable point is $m=0$.
From this perspective opinion dynamics can be understood as a movement of a public opinion $m$ in a symmetric bistable
potential $V(m)$. Therefore we can expect random transitions between the neighboring potential
wells, at least in a case of finite systems. 

\section{Analytical results for the finite system}
To investigate spontaneous transition between two metastable states, let us now derive the evolution equation for the finite system. Again, we start from calculating the probabilities $p_+$ that the magnetization increases $m \rightarrow m+\Delta_N$ and $p_-$ that magnetization decreases $m \rightarrow m-\Delta_N$. 
In the case of the finite system, probabilities of choosing active configurations are the following:
\begin{eqnarray}
p_{\uparrow\uparrow\downarrow} & = & \frac{N_{\uparrow}}{N} \cdotp \frac{N_{\uparrow}-1}{N-1} \cdotp \frac{N_{\downarrow}}{N-2}, \nonumber \\
p_{\uparrow\uparrow\uparrow} & = & \frac{N_{\uparrow}}{N} \cdotp \frac{N_{\uparrow}-1}{N-1} \cdotp \frac{N_{\uparrow}-2}{N-2}, \nonumber \\
p_{\downarrow\downarrow\uparrow} & = & \frac{N_{\downarrow}}{N} \cdotp \frac{N_{\downarrow}-1}{N-1} \cdotp \frac{N_{\uparrow}}{N-2}, \nonumber \\
p_{\downarrow\downarrow\downarrow} & = & \frac{N_{\downarrow}}{N} \cdotp \frac{N_{\downarrow}-1}{N-1} \cdotp \frac{N_{\downarrow}-2}{N-2}.  
\end{eqnarray}
Again, using (\ref{eq:prob}) transition probabilities can be calculated and finally:
\begin{eqnarray}
p_+ & = & \frac{p_1N_{\uparrow}(N_{\uparrow}-1)N_{\downarrow} + p_2N_{\downarrow}(N_{\downarrow}-1)(N_{\downarrow}-2)}{N(N-1)(N-2)}, \nonumber \\
p_- & = & \frac{p_1N_{\downarrow}(N_{\downarrow}-1)N_{\uparrow} + p_2N_{\uparrow}(N_{\uparrow}-1)(N_{\uparrow}-2)}{N(N-1)(N-2)}, \nonumber \\
p_0 & = & 1-(p_++p_-).
\label{eq:prob1} 
\end{eqnarray}
Using formulas (\ref{eq_NpNm}) we obtain:
\begin{equation}
N_{\uparrow} = \frac{1+m}{2}N,~~~N_{\downarrow} = \frac{1-m}{2}N,
\end{equation}
and therefore we can rewrite transition probabilities $p_+,p_-$ (\ref{eq:prob1}) as functions of $m$.

Let $P(m,t)$ denotes the probability density function of the magnetization $m$ at time $t$. Time evolution of $P(m,t)$ is described as usual by the master equation:
\begin{eqnarray}
P(m,t+\Delta_t) &= p_{+}(m-\Delta_N)P(m-\Delta_N,t) \nonumber\\
                &+ p_{-}(m+\Delta_N)P(m+\Delta_N,t) \nonumber\\
                &+[1-p_{+}(m)-p_{-}(m)] P(m,t),
\label{eq_evol}
\end{eqnarray}
where $\Delta_t=1/N,\Delta_N=2/N$.

Evolution equation allows to calculate numerically not only the asymptotic states, but also the time evolution of the system, probabilities of transitions between two asymptotic stable states and probability density function (PDF) of waiting times (see the next section).

For large, but finite systems ($1<<N<\infty$) we can solve the master equation analytically. Let us first rewrite it in the following form: 
\begin{eqnarray}
&P(m,t+\Delta_t)- P(m,t) = \nonumber \\
&=[\rho(m-\Delta_N)P(m-\Delta_N,t)+\rho(m+\Delta_N)P(m+\Delta_N,t) - 2\rho(m)P(m,t)] \nonumber\\
&-[F(m-\Delta_N)P(m-\Delta_N,t)-F(m+\Delta_N)P(m+\Delta_N,t)]/2,
\end{eqnarray}
where:
\begin{eqnarray}
\rho &=&\frac{p_{+}+p{-}}{2} 
\end{eqnarray}
and $F=p_{+}-p_{-}$ is an effective force defined already in Eq. (\ref{effective_force}).

For $1<<N<\infty$ above equation can be approximated by the following differential equation:
\begin{eqnarray} 
\frac{1}{N}\frac{\partial }{\partial t} P(m,t) = \frac{1}{N^2} \frac{\partial^2}{\partial m^2} (\rho P(m,t)) 
- \frac{1}{N} \frac{\partial }{\partial m} (FP(m,t)),
\end{eqnarray}
which is a well known Fokker-Planck equation with diffusion coefficient:
\begin{equation}
\rho = \frac{p_{+}+p{-}}{2} = \frac{p_1+p_2}{8}+\frac{3p_2-p_1}{8}m^2
\end{equation}
and drift 
\begin{equation}
F = p_+-p_-=- \left(\frac{3p_2-p_1}{4}+\frac{p_1+p_2 }{4}m^2\right) m.
\end{equation}
The stationary equation takes the following form:
\begin{equation}
\frac{1}{\rho}\frac{\partial}{\partial m} (\rho P(m))  =  \frac{NF}{2\rho} 
\end{equation}
which has a general solution:
\begin{equation}
P(m) = \frac{C}{\rho} \exp \int \frac{N F}{2\rho} dm,
\end{equation}
where $C$ is normalizing constant.
Above equation has two qualitatively different stationary solutions depending on the quantity $d$ defined by (\ref{def_d}):
\begin{itemize}
\item For $d=0$ we obtain:
\begin{equation}
P(m) = C \exp ( - N m^4/4 ), 
\end{equation}
\item For $d > 0$:
\begin{equation}
P(m) = C (1-d m^2) ^{-1}(1-d m^2)^{(d^{-2}-1)N/2}\exp(N m^2/2d).
\label{pm_dne0}
\end{equation}
\end{itemize}
$P(m)$ has two maxima at $m_{max}=m_{\pm}=\pm \sqrt{d}$. It should be noticed here that $m_{max}$ is equal to $m_{st}$ found in (\ref{m_st}) for the infinite system.

\begin{figure}
\begin{center}
\includegraphics[width=12cm]{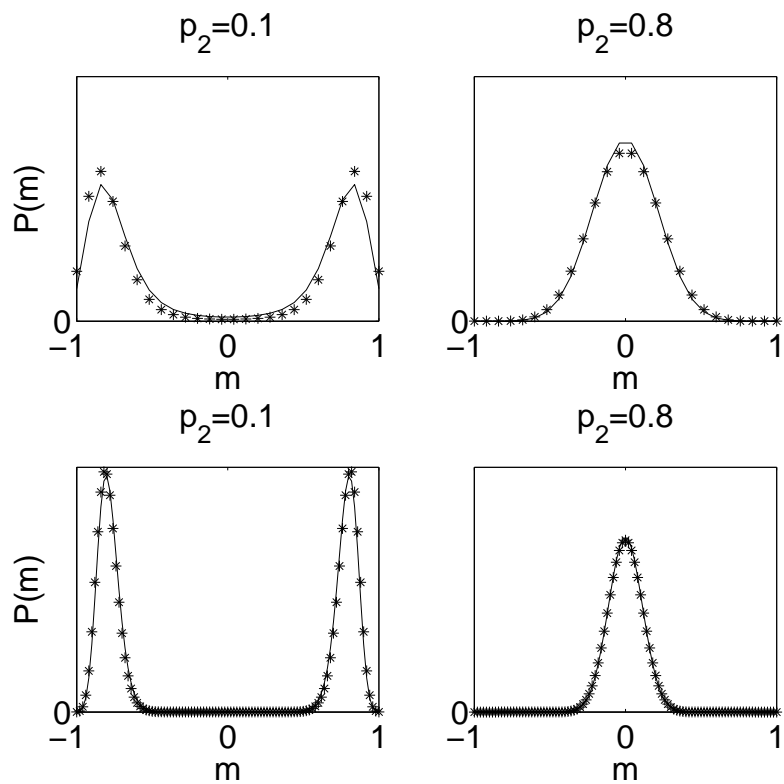}
\caption{Stationary density probability function of magnetization from analytical formula (\ref{pm_dne0}) (solid line) and obtained numerically directly from the master equation (\ref{eq_evol}) (stars) for $N=25$ (upper panel) and $N=100$ (bottom panel); $p_1=1$.}
\label{comp_analit}
\end{center}
\end{figure}

We would like to stress here that above analytical formulas for the stationary density probability function of magnetization $P(m)$ have been derived under assumption that the system size is large i.e. $N>>1$. This allowed us to replace discrete by continues differential equation. However, obtained results agree very well with exact results, that can be obtained numerically from the evolution equation (\ref{eq_evol}), already for $N=100$ (see Fig. \ref{comp_analit}). 

In the next section we compare our analytical and numerical results with Monte Carlo simulations and first of all investigate random transitions between two stable states $m_{st}=m_{\pm}$, that can be observed in a case of a finite system.

\section{Monte Carlo Simulations and comparison with analytical results}
We have performed computer simulations on a complete graph for $p_2 \in (0,1)$ for several system sizes -- from $N=50$ to $200$. As we have noticed in the previous section asymptotic behavior of our system is determined only by the ratio $r=p_2/p_1$ between anticonformity and conformity. From social point of view the probability $p_1$ of conformal behavior  is always greater then the probability $p_2$ of anticonformal behavior, therefore we have decided to choose $p_1=1$ and $p_2 \in [0,1]$ (the same values of parameters $p_1,p_2$ has been chosen in \cite{KSW2010}). That means that for $p_2=0$ we have original Sznajd model. 

\begin{figure}
\begin{center}
\includegraphics[width=12cm]{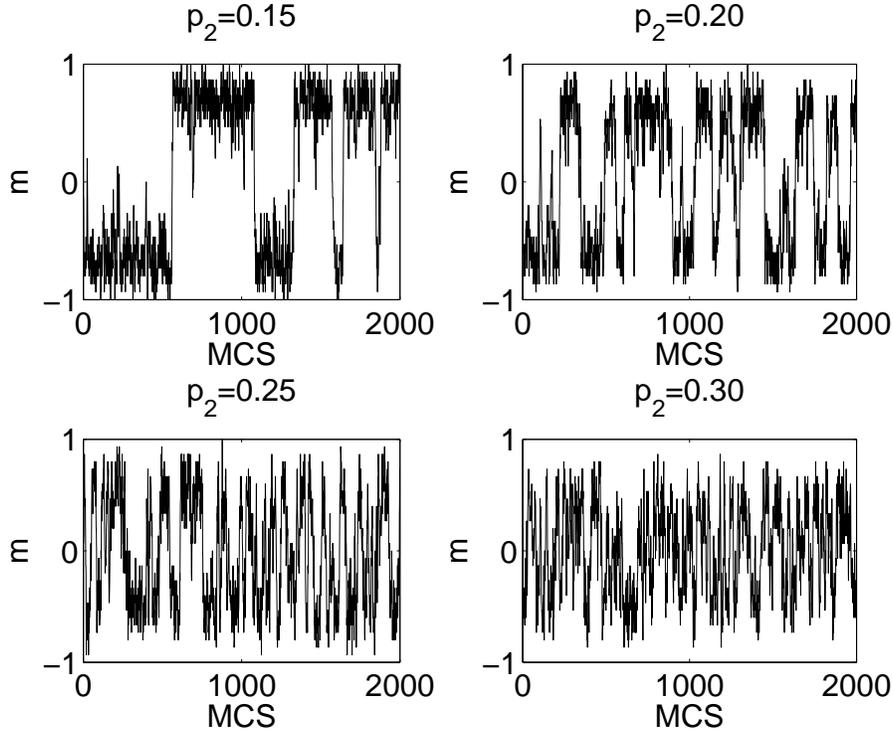}
\caption{Time evolution of the public opinion for several values of anticonformity level $p_2$ and $p_1=1$ -- for the low level of anticonformity spontaneous transition between two stable states occur. This behavior reminds a typical stochastic realization sampling the random waiting times in a symmetric bistable potential, with stable states located at $m=m_{\pm}$ \cite{HTB1990}.}
\label{mt}
\end{center}
\end{figure}

The time evolution of the public opinion, shown in Figure \ref{mt}, reminds a typical stochastic realization sampling the random waiting times in a symmetric bistable potential, with stable states located at $m=m_{\pm}$ \cite{HTB1990}. This is seen that values $m_{\pm}$ are $p_2$-dependent and above certain threshold $p_2>p_2^*$ system becomes monostable. For $p_2 \rightarrow 0$ transitions between two stable state are rare, i.e. stability of states $m=m_{\pm}$ is high. With increasing aniconformity level $p_2$ transitions are more and more frequent -- stability of $m=m_{\pm}$ decreases. This behavior is clearly seen from the dependence between PDF of public opinion and $p_2$, which can be obtained from Monte Carlo (MC) simulations as well as analytical treatment presented in the previous section.

\begin{figure}
\begin{center}
\includegraphics[width=12cm]{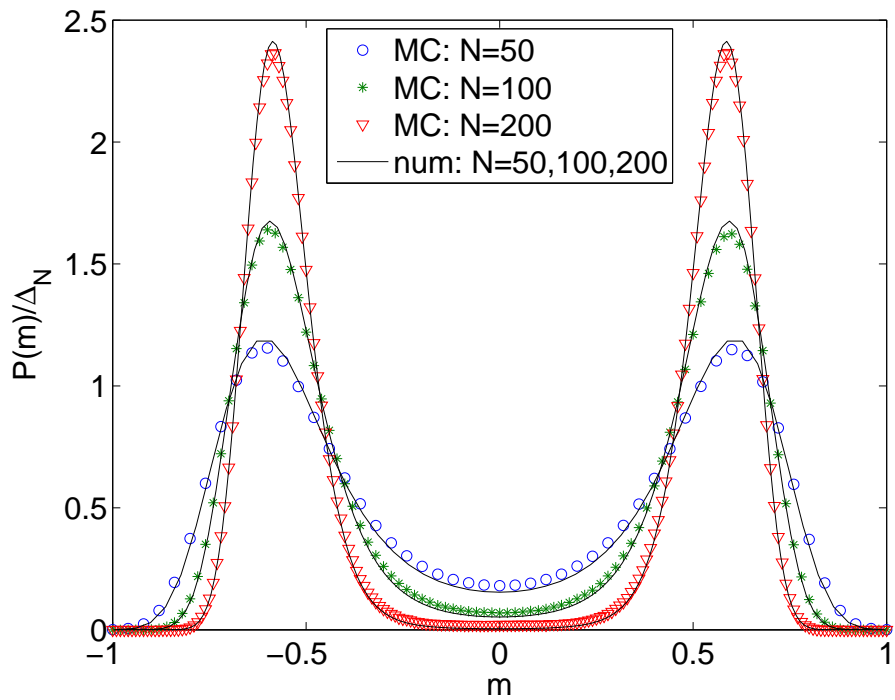}
\caption{Probability density function of the public opinion $P(m)$ divided by $\Delta_N=2/N$ for several lattice sizes and anticonformity level $p_2=0.2$. Numerical results obtained from equation (\ref{eq_evol}) agree with Monte Carlo (MC) simulations.}
\label{P-m}
\end{center}
\end{figure}

\begin{figure}
\begin{center}
\includegraphics[width=12cm]{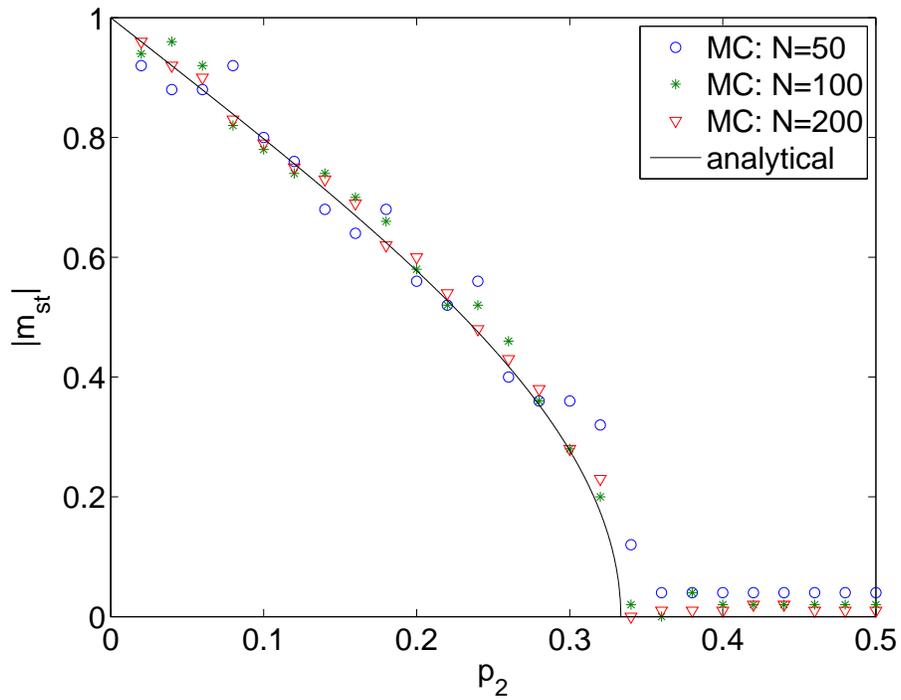}
\caption{Asymptotic absolute value of a public opinion as a function of anticonformity level $p_2$. Continuous phase transition is clearly visible for $p_2=p^* = 1/3$. Below this value there are two possible stable states. Due to fluctuations, spontaneous transitions between these two states are possible. Above the transition point $p^*$ there is no majority in the system. From the social point of view, community is in a stalemate state. Analytical result obtained for the infinite system agrees with Monte Carlo simulations.}
\label{pdf}
\end{center}
\end{figure}

In Figure \ref{P-m} PDFs of public opinion for $p_2=0.2$ for several lattice sizes are presented. As seen, numerical results obtained from (\ref{eq_evol}) agree with Monte Carlo simulations, which is expected since we deal with a complete graph. 
In the next figure \ref{P-m-6} we present probability density function of the public opinion $P(m)$ depending on anticonformity level $p_2$. We have shown only analytical results to make figure more legible but MC simulations give consistent results for the whole range of anticonformity level $p_2 \in (0,1)$.  
\begin{figure}
\begin{center}
\includegraphics[width=12cm]{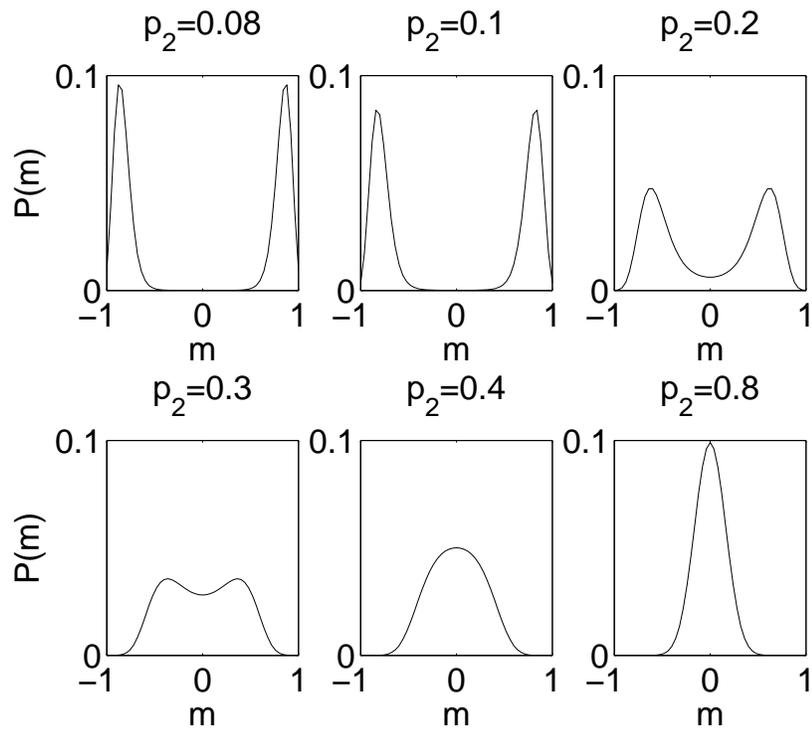}
\caption{Probability density function of the public opinion $P(m)$ for $p_1=1$ calculated numerically from Eq. (\ref{eq_evol}).}
\label{P-m-6}
\end{center}
\end{figure}
As seen from figure \ref{P-m-6} for small values of $p_2$ the system is bistable with modes at $m=m_{\pm}$. With increasing $p_2$ the distance and the hight of the separation between modes decreases. For certain threshold value $p_2=p_2^*$ both modes approach $m=0$ and PDF of public opinion $P(m)$ becomes unimodal. If we choose as an order parameter asymptotic absolute value of public opinion, we see that the system exhibits continuous phase transition at $p_2=p_2^* = 1/3$ (see Fig. \ref{pdf}), as predicted analytically. Below the critical value of anticonformity two symmetric stable states exist. Due to fluctuations, spontaneous transitions between these two states are possible. Above the transition point $p^*$ there is no majority in the system, i.e. $m$ fluctuates around zero. From the social point of view, community is in a stalemate state. It should be recall here that in the case of one-dimensional system we have also observed spontaneous transitions between two stable states. However, on contrary to the case of a complete graph, stable states were equal $m = \pm 1$ for any value of $p_2<p_2^*$. Results on a complete graph are more reasonable for social systems -- majority instead of unanimity is present for low level of anticonformity. Moreover, the level of the majority depends on $p_2$. 
Similar results regarding the type of the phase transition and it's dependence on the topology were previously obtained by de la Lama et al., but spontaneous transitions between modes below the critical point were not considered in \cite{LLW2005}.  

\begin{figure}
\begin{center}
\includegraphics[width=12cm]{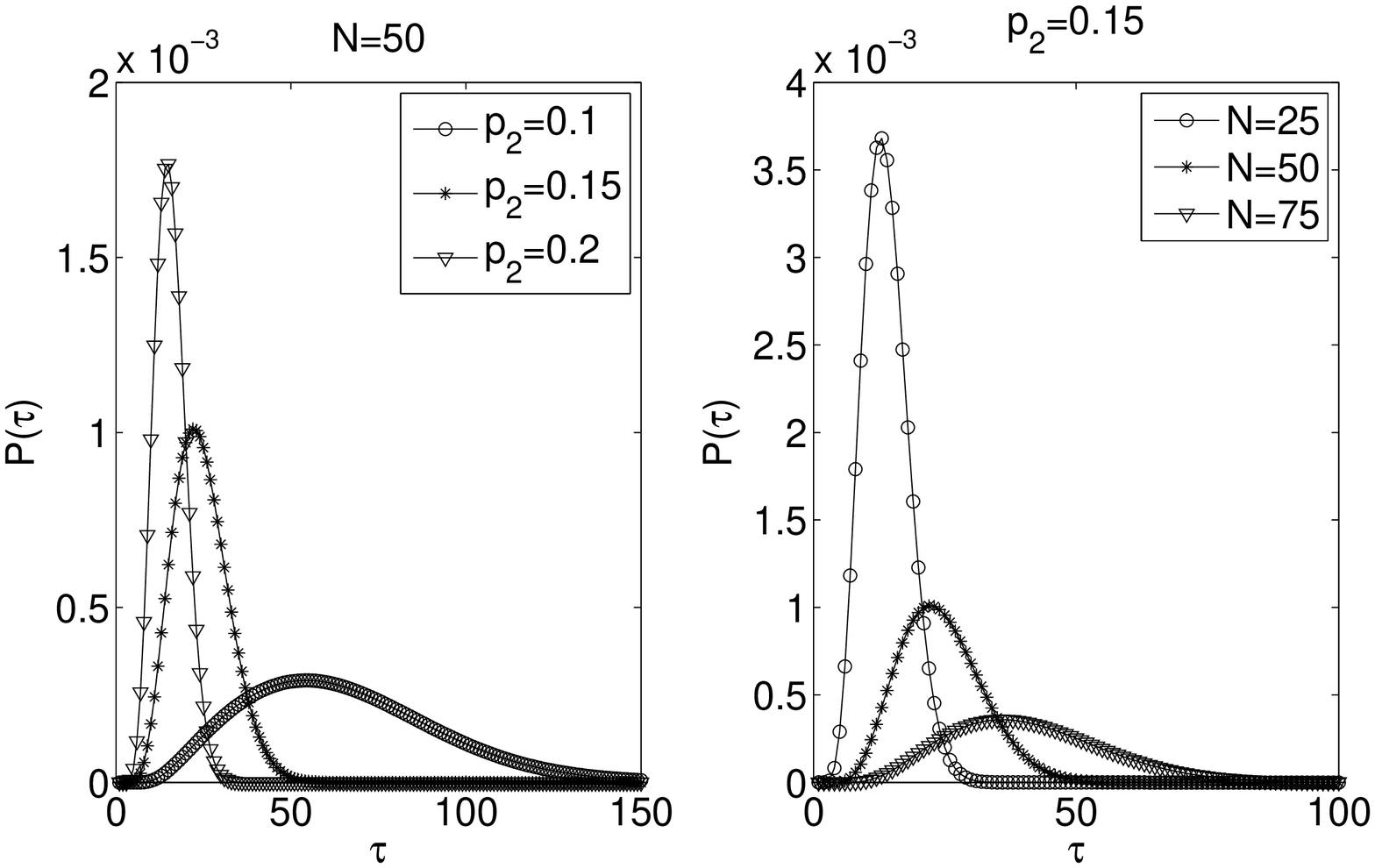}
\caption{Probability density function of waiting times in stable states for several values of anticonformity level $p_2$ (left panel) and the size $N$ (right panel).}
\label{transPrado}
\end{center}
\end{figure}

\begin{figure}
\begin{center}
\includegraphics[width=12cm]{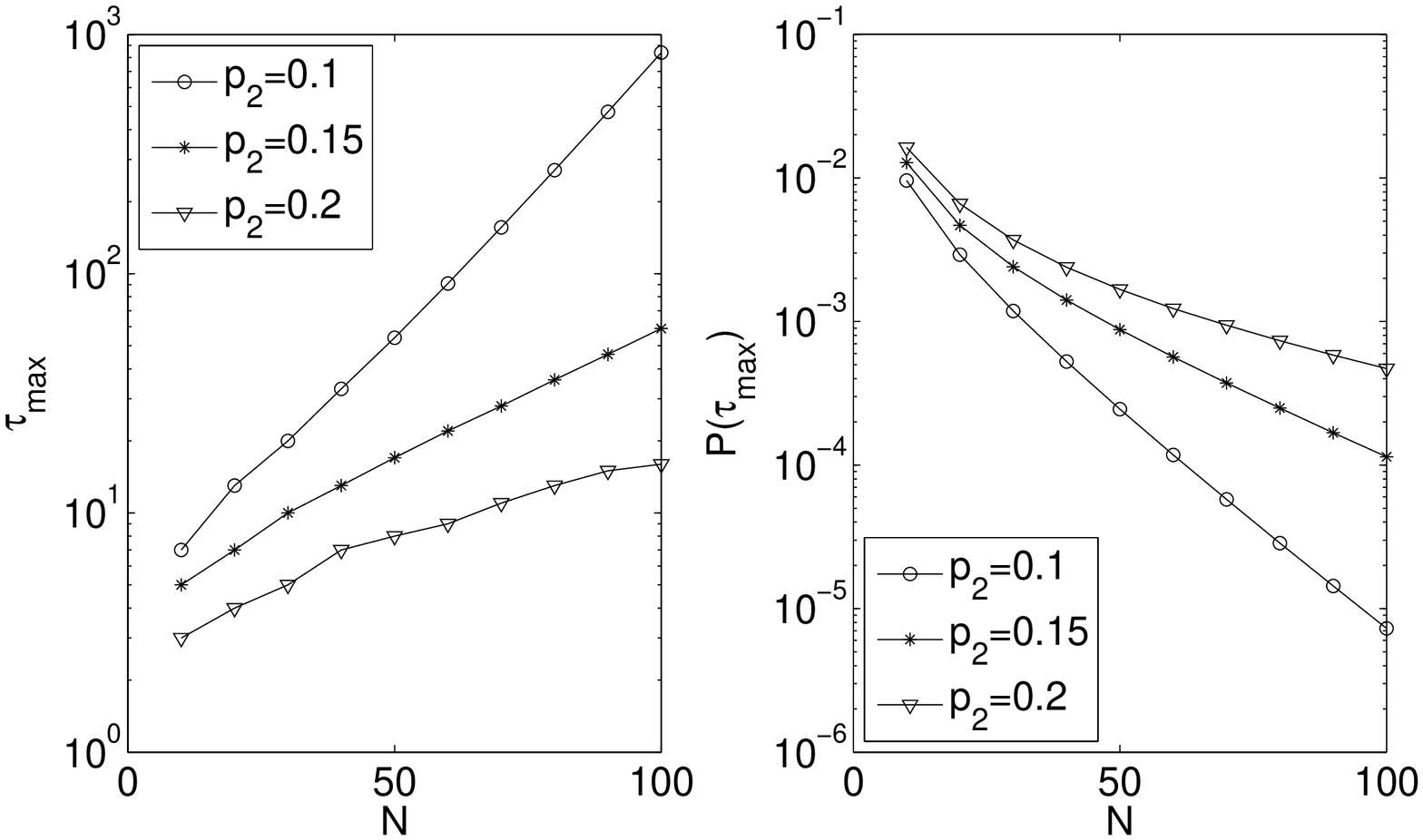}
\caption{Dependence between the system size $N$ and the most probable waiting time $\tau{_{max}}$ is shown in the left panel. In the right panel -- dependence between the system size $N$ and the probability $P(\tau{_{max}})$. As expected the typical waiting time increases with the system size and decreases with $p_2$.}
\label{transMax}
\end{center}
\end{figure}

Let us now focus on those transition and measure the waiting time $\tau$ that system spends in one of two stable states before jumping to the second one. We expect that waiting time $\tau$ increases with the system size and decreases with $p_2$. From Eq. \ref{eq_evol} we are able to calculate the probability density function of waiting times. It occurs that the typical, i.e most probable waiting time exists (see Fig. \ref{transPrado}). As expected the typical waiting time increases with the system size and decreases with $p_2$. For low values of anticonformity level $p_2$ the most probable transition time $\tau{_{max}}$ grows almost exponentially with the system size $N$, which indicates that in the infinite system spontaneous transitions never occur (see Fig. \ref{transMax}). 

\section{Summary}
In this paper we have investigated a simple model of opinion dynamics with contrarians on a complete graph. Using Monte Carlo simulations and analytical calculations we have shown that below the critical level of anticonformity the system is bistable spending most of the time in one two possible stable states. Above the critical value system becomes monostable and no majority exists in a system - so called stalemate situation. On contrary to one-dimensional case the phase transition is continuous.  From social perspective,  especially interesting results regard the spontaneous transitions between stable/metastable states. As shown here using analytical treatment, in a case of finite system spontaneous transitions between states can be observed and the typical waiting time exists.  

Random transitions between the neighboring potential wells caused by fluctuational forces were observed in various type of systems \cite{HTB1990}. However, most of the papers concentrate on so called stochastic resonance -- the response of s bistable system to a periodic external force in the presence of noise \cite{GHJM2009,GHJM1998,WLL2006}. Yet in many nonlinear systems coherent transitions are not stimulated by an external force \cite{GDNH1993} and this is also probably the case of social systems.

\end{document}